\documentclass[prb,twocolumn,floatfix,showpacs]{revtex4-1}
\usepackage{graphicx,color}
\usepackage{amsmath,amssymb,bm}
\DeclareMathOperator{\sgn}{sgn}

\newcommand{\la}{\Lambda}
\newcommand{\tn}{\textnormal}
\newcommand{\cpra}[3]{Phys.~Rev.~A {\bf #1}, #2 (#3)}
\newcommand{\cprb}[3]{Phys.~Rev.~B {\bf #1}, #2 (#3)}
\newcommand{\cprl}[3]{Phys.~Rev.~Lett.~{\bf #1}, #2 (#3)}

\newcommand{\cbook}[2]{\textit{#1} (#2)}

\definecolor{darkred}{rgb}{0.90,0,0}
\definecolor{darkgreen}{rgb}{0,0.60,.2}
\definecolor{darkblue}{rgb}{0,0,1}
\definecolor{grey}{cmyk}{0,0,0,0.25}
\definecolor{orange}{cmyk}{0,0.6,0.8,0}

\begin{document}

\title{Approaching Many-Body Localization from Disordered Luttinger Liquids\\ via the Functional Renormalization Group}

\author{C.\ Karrasch}
\author{J.\ E.\ Moore}

\affiliation{Department of Physics, University of California, Berkeley, California 95720, USA}
\affiliation{Materials Sciences Division, Lawrence Berkeley National Laboratory, Berkeley, CA 94720, USA}

\begin{abstract}

We study the interplay of interactions and disorder in a one-dimensional fermion lattice coupled adiabatically to infinite reservoirs. We employ both the functional renormalization group (FRG) as well as matrix product state techniques, which serve as an accurate benchmark for small systems. Using the FRG, we compute the length- and temperature-dependence of the conductance averaged over $10^4$ samples for lattices as large as $10^{5}$ sites. We identify regimes in which non-ohmic power law behavior can be observed and demonstrate that the corresponding exponents can be understood by adapting earlier predictions obtained perturbatively for disordered Luttinger liquids. In presence of both disorder and isolated impurities, the conductance has a universal single-parameter scaling form. This lays the groundwork for an application of the functional renormalization group to the realm of many-body localization.

\end{abstract}

\maketitle



\section{Introduction}

It has been known since Anderson's work in 1958 that disorder can localize the eigenstates of a non-interacting system. \cite{anderson} This single-particle localization physics is now well understood theoretically and was observed experimentally (for reviews see Refs.~\onlinecite{andrev1,andrev2,andrev3}). In one or two spatial dimensions, an arbitrarily small amount of disorder will localize any eigenstate in the spectrum, but in 3d a so-called mobility edge can exist which separates localized states at the lower end of the spectrum from extended states at high energies. The transition, which can, e.g., be triggered by varying the disorder strength, is the so-called Anderson transition.

The notion of localization heavily relies on a single-particle picture. One might generally expect that upon adding interactions (i.e., collisions), every state gets delocalized. However, in 2006, Basko, Aleiner, and Altshuler suggested \cite{mbl0} that the localized phase can exist even in presence of interactions and that a finite temperature phase transition can occur between phases with zero and finite conductivity. This phase transition is not a thermodynamic (equilibrium) transition but a dynamical quantum phase transition which takes place on the level of the many-body eigenstates and is beyond standard Mermin-Wagner arguments. For one-dimensional lattice systems, the stability of localized states towards adding interactions -- i.e., the existence of a `many-body localized' phase -- has subsequently been established fairly convincingly by a number of numerical \cite{mblhuse1,mblprosen,mbled1,mblhuse2,mblarea,mbljens2,edluitz} and analytical \cite{mbltop1,mbltop2,mblproof} studies. Moreover, there is evidence that a transition into a delocalized phase occurs if the ratio between the interaction and the disorder strength is increased. \cite{mblhuse2,mbljens2,edluitz} Physical properties have been investigated partially.\cite{mbljens,mbltrans1,mbltrans2,mbltrans3,mblbath1,mblbath2}

To date, the world of many-body localization (MBL) has primarily been explored numerically by exact diagonalization, because many more advanced tools have difficulty in resolving excited-state properties and transport. Techniques such as density matrix renormalization group \cite{dmrgrev} often increase the accessible system size only slightly. In order to deepen our understanding of MBL physics, it would be desirable to employ different methods which are complementary in their strengths and shortcomings. In this paper, we propose the functional renormalization group (FRG),\cite{frgrev2} which formulates an a priori exact RG flow on the level of Green functions, as one such method. The FRG is capable of studying large systems but is approximate since in practice, the infinite hierarchy of FRG equations has to be truncated. This raises the question of how well FRG calculations describe transport in systems with both interactions and disorder. It is one goal of our work to investigate this issue.

In the realm of MBL, the starting point to understand the interplay of disorder and correlations is the conventional (non-interacting) Anderson insulator. In one dimension, however, this problem was first tackled in the opposite limit of weak disorder being added to a clean correlated system. In absence of disorder, interacting 1d systems generically feature low-energy excitations which are not fermionic quasiparticles but collective (bosonic) modes; as a consequence, their correlation functions exhibit anomalous power laws in space and time with interaction-dependent exponents. \cite{voit,giamarchi,kurtrev} This so-called Luttinger liquid (LL) physics is often described using the exactly-solvable Tomonaga-Luttinger (TL) model,\cite{tomonaga,luttinger} which is then argued to be the fixed point model governing the physics of a large class of 1d systems at low energies.\cite{haldane}

The physics of Luttinger liquids in presence of isolated impurities \cite{lutherpeschel,mattis,kf} or weak disorder \cite{lutherpeschel2,apel1,apel2,apel3,gs1,gs2} was studied extensively using field-theoretical versions of the TL model. Power law behavior of physical quantities can be understood, e.g., from scaling dimensions computed perturbatively around the Luttinger liquid fixed point. However, important questions remained: (1) Can one justify some of the approximations (e.g., about the way disorder is treated) made in these calculations by employing a different methodology -- i.e., by making different approximations? (2) Can one verify the power laws directly for a microscopic lattice without having to resort to the usual arguments necessary to show that the field theory indeed describes its low-energy physics? (3) What is the temperature- or length scale (such as the localization length) below/above which these power laws can be observed for a given model? (4) What is the physics on all scales? These questions were addressed in details in the case of isolated impurities (see, e.g., the discussion in Refs.~\onlinecite{LLvolker,LLvolker2,LLvolker2a,LLvolker3}) but in comparison only sparsely in presence of disorder.\cite{eckern,ps,schuster,mackinnon,laflorencie,mirlin1,mirlin2,weiss,kafri}

We propose the functional renormalization group (FRG) \cite{frg0a,frg0b,frgrev1,frgrev2} as an alternative method to study transport in interacting, disordered systems. The key drawback of the FRG is that in practice its flow equation hierarchy needs to be truncated via an expansion w.r.t.~the two-particle interaction on the right-hand side. Hence, it is imperative to benchmark the capabilities of low-order approximations; we resort to the density matrix renormalization group (DMRG), which yields accurate results but is limited to small systems, as a frame of reference. On the upside, (a) the FRG flow incorporates single-particle disorder exactly, makes no assumptions about the existence or absence of intermediate fixed points between high and low energies, and can often be continued to zero cutoff, (b) the FRG can be used on the Matsubara axis (equilibrium) or in real-time Keldysh space,\cite{kelfrg,tfrg} and it is not a low-entanglement approximation, (c) one can treat both open and close systems that are generically much larger than those accessible via exact diagonalization, and (d) one can describe the physics on all scales.

We start by analyzing a size-$L$ lattice of spinless fermions with repulsive interactions $U$ in presence of weak disorder $\eta\ll1$. The field-theoretical studies of disordered Luttinger liquids predict power laws whose exponents contain the interaction $U$ to linear order; hence, a first-order FRG approximation is a reasonable starting point to study the physics in this limit. We adiabatically couple the system to reservoirs and compute the length- and temperature-dependence of the conductance on all scales. First, the FRG results are compared with DMRG data for lattices of $O(10)$ sites. For larger $L$, we observe non-ohmic power laws and eventually exponential decays. This is consistent with the system being localized. The interaction- and disorder-dependence of the localization length is documented. We demonstrate that in presence of both isolated impurities $V$ as well as weak disorder, the conductance $G(L)$ shows a crossover between two power laws, and its values for different $L$, $V$, and $\eta$ can be collapsed on a single universal curve.

We find that even for quite strong interactions our FRG calculations are in very good agreement with the DMRG data as well as with the theoretical predictions based on disordered Luttinger liquids. Many-body localization with zero conductivity at nonzero temperature is expected to appear as a conductance that, at fixed temperature, decreases exponentially with the length of the disordered region -- similarly to the zero-temperature behavior of a standard localized system, albeit with a possibly different localization length. At intermediate disorder strengths, there is a regime at nonzero temperature where the exponential scaling of conductance persists consistent with MBL -- i.e., FRG can effectively give an upper bound on the conductivity in the thermodynamic limit, while not proving it is strictly zero. We find evidence for a crossover into a metallic phase for attractive interactions. A more systematic study of the limit of strong disorder and many-body localization, which is a strong-coupling phenomenon, requires a full second-order truncation scheme. Our present work lays the groundwork for this calculation (which itself is beyond the scope of this paper) and for a description of the full crossover between the limits of weak and strong disorder. While it is not garantueed that a second-order scheme would succeed in describing MBL quantitatively, one would expect that the MBL phase can be detected and that some of its features can be analyzed qualitatively from the second-order flow to strong coupling.

\section{Model}
\label{sec:model}

We consider one-dimensional spinless fermions living on a lattice of size $\tilde L=L+2L_c$:
\begin{equation}\label{eq:hll}\begin{split}
H_\tn{LL} = \sum_{l=1}^{\tilde L-1}\Big(-t_l c_l^\dagger &c_{l+1}^{\phantom{\dagger}} + \tn{h.c.} + V_ln_l + U_l\tilde n_l\tilde n_{l+1}\Big),
\end{split}\end{equation}
where $n_l=c_l^\dagger c_{l}^{\phantom{\dagger}}$, $\tilde n_l=n_l-1/2$, and $t_l$ denote the nearest-neighbor hopping amplitudes. We will mainly set $t_l=t$ and model disorder via random potentials $V_l$. In order to adiabatically connect the system to reservoirs, we switch on the Coulomb interaction smoothly over a few lattice sites $L_c\ll L$ (we will comment on the values $L_c$ and $s$ below):
\begin{equation}\label{eq:usmooth}
U_{l\leq\tilde L/2} = U
\begin{cases}
\Big[\frac{\pi}{2}+\arctan\big[s(l-1-L_c/2)\big]\Big] & l \leq L_c \\
1 & l > L_c\,,
\end{cases}
\end{equation}
and $U_{\tilde L-l}=U_{l}$. Finally, the coupling to left (L) and right (R) Fermi liquid leads is described by
\begin{equation}\begin{split}
H_\tn{lead} = \sum_{k} \big[\epsilon_{k} & f_{L,k}^\dagger f_{L,k}^{\phantom{\dagger}} + \epsilon_{k} f_{R,k}^\dagger f_{R,k}^{\phantom{\dagger}}\\
 + &\tau \big(f_{L,k}^\dagger c_1  + f_{R,k}^\dagger c_{\tilde L} +\tn{h.c.}\big)\big]\,.
\end{split}\end{equation}
Prominent choices for the dispersion $\epsilon_k$ are i) tight-binding leads governed by the Hamiltonian of Eq.~(\ref{eq:hll}) with $V_l=U_l=0$, and ii) the wide-band limit of structureless reservoirs described by a single hybridization energy $\Gamma=\pi\rho_\tn{lead}(0)\tau^2$, where $\rho_\tn{lead}(\omega)$ is the local density of states at the chemical potential. We have checked explicitly that both yield the same results in the low-energy limit.

\section{Method}
\label{sec:method}

\begin{figure}[t]
\includegraphics[height=1.0cm,clip]{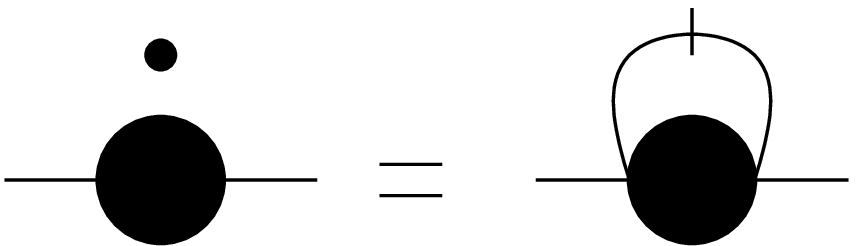}\\
\includegraphics[height=1.0cm,clip]{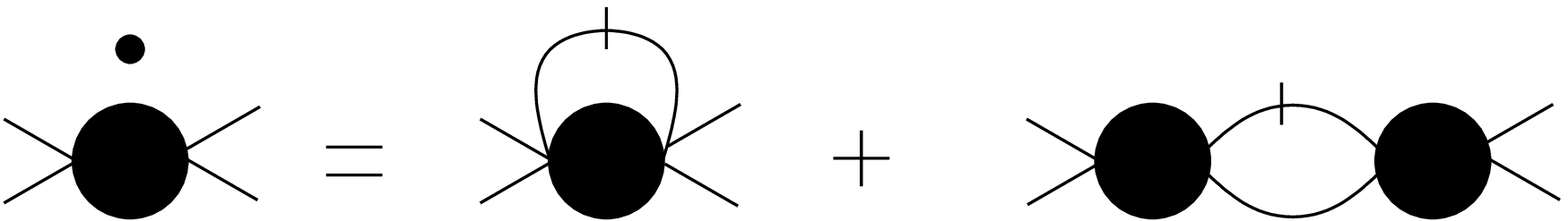}
\caption{(Color online) Schematic representation of the flow equations for the self-energy and the effective two-particle scattering (a $n$-particle vertex has $2n$ external legs).  }
\label{fig:floweq}
\end{figure}

\subsection{Functional Renormalization Group}

The functional renormalization group is one implementation of Wilson's general RG idea for interacting many-particle systems.\cite{frgrev1,frgrev2} It starts with introducing an energy cutoff $\la$ into the non-interacting Matsubara Green function $G^0$ of the system under consideration (note that the method can also be set up on the Keldysh axis). We choose a multiplicative infrared cutoff in Matsubara frequency space, which at zero temperature takes the simple form $\Theta(|i\omega|-\la)$; the finite-$T$ analogue can be found in Ref.~\onlinecite{LLvolker3}. We consider the flow of many-particle vertex functions, the lowest of which are the self-energy and the effective two-particle scattering. By virtue of the replacement
\begin{equation}
 G^0(i\omega) \to G^{0,\la}(i\omega) = \Theta(|i\omega|-\la) G^0(i\omega)\,,
\end{equation}
every such vertex function acquires a $\la$-dependence. If one takes the derivative with respect to $\la$ (which can, e.g., be accomplished using generating functionals), one obtains an infinite hierarchy of flow equations that can be represented diagrammatically (see Fig.~\ref{fig:floweq}); a detailed derivation can be found in Ref.~\onlinecite{frgrev2}. Subsequent integration from $\la=\infty$ down to the cutoff-free system $\la=0$ leads to an in principle exact solution of the many-particle problem. In practice, the infinite hierarchy needs to be truncated, rendering the FRG an approximate method. The most simple truncation scheme is to neglect the flow of the two-particle vertex by setting it to its initial value (the bare Coulomb interaction $U$) in the self-energy flow equation. This approximation is strictly correct only to leading order in $U$ but contains an infinite resummation of Feynman diagramms (since the self-energy feeds back into its own flow). Similarly, the second-order truncation scheme is obtained by setting the three-particle vertex to its initial value (zero), which yields a closed set of flow equations for both the two-particle vertex and the self-energy.

In this paper, we consider the flow of the self-energy and partially incorporate second-order contributions by parametrizing the two-particle vertex as purely local and energy-independent, i.e., as effective on-site interactions $U_l^\la$.\cite{commentflow} It turns out that accounting for the flow of $U_l^\la$ renders the higher-order contributions to Luttinger liquid exponents associated with isolated impurities more accurate (the exponents associated with disorder do not improve significantly). We emphasize that this is a purely pragmatic approach - the resulting approximation is still exact only to first order. The flow of the self-energy can by can be expressed in terms of effective hoppings $t_l^\la$ and on-site energies $V_l^\la$; the flow equations explicitly read
\begin{equation}\label{eq:flow}\begin{split}
\partial_\lambda V_l^\la &= -\frac{1}{\pi}\tn{Re}\,\big[U_{l-1}^\la \tilde G_{l-1,l-1}^\la(i\omega_n^\la)+U_{l}^\la \tilde G_{l+1,l+1}^\la(i\omega_n^\la)\big] \\
\partial_\lambda t_l^\la & = -\frac{1}{\pi}\tn{Re}\,\big[U_l^\la \tilde G_{l,l+1}^\la(i\omega_n^\la)\big] \\
\partial_\lambda U_l^\la & =\frac{U_l^\la}{\pi}\Big\{ 2 U_l^\la \big[\tn{Re}\, \tilde G_{l,l}^\la(i\omega_n^\la)\tn{Re}\, \tilde G_{l+1,l+1}^\la(i\omega_n^\la)\\
&\hspace*{1.65cm}  - \tn{Re}\, \tilde G_{l,l+1}^\la(i\omega_n^\la)\tn{Re}\, \tilde G_{l,l+1}^\la(i\omega_n^\la)\big]\\
& - U_{l-1}^\la \tn{Re}\,\tilde G_{l-1,l}^\la(i\omega_n^\la)^2
- U_{l+1}^\la \tn{Re}\,\tilde G_{l,l+1}^\la(i\omega_n^\la)^2 \Big\}\,.
\end{split}\end{equation}
The initial conditions are given by $V_l^{\la\to\infty}=V_l$, $t_l^{\la\to\infty}=t_l$, and $U_l^{\la\to\infty}=U_l$. Boundary conditions are formally imposed by setting $U^\la_{-1}=U^\la_{\tilde L}=0$. $\tilde G^\la(i\omega)$ denotes the flowing single-particle Matsubara Green function, and $\omega_n^\la$ is the Matsubara frequency closest to $\la$. The non-interacting leads can be `projected out' analytically via equation-of-motion techniques, and the calculation of $\tilde G^\la(i\omega)$ then reduces to the inversion of a $\tilde L\times\tilde L$ matrix defined by 
\begin{equation}\begin{split}
& \big[\tilde G^\la(i\omega)^{-1}\big]_{l,l} = i\omega - V_l^\la -\tau^2 g_\tn{lead}(i\omega)(\delta_{l,1}+\delta_{l,\tilde L}) \\
& \big[\tilde G^\la(i\omega)^{-1}\big]_{l,l+1} = \big[\tilde G^\la(i\omega)^{-1}\big]_{l+1,l} = t_l^\la\,,
\end{split}\end{equation}
where $g_\tn{lead}(i\omega)$ is the local Green function of an isolated lead [which in the wide-band limit is determined by $\tau^2g_\tn{lead}(i\omega)= -i\Gamma\sgn(\omega)$]. Due to the tri-diagonal structure, this inversion can be carried out with a computational effort scaling linearly with $\tilde L$.\cite{LLvolker2a} The flow equations can be integrated using standard Runge-Kutta routines. Finally, one obtains the conductance (in units of $e^2/h=1$) from
\begin{equation}\label{eq:cond}\begin{split}
G(L,T) = -4\pi^2\tau^4\int & d\omega \Big[f'(\omega)\rho_\tn{lead}(\omega)^2\\\times &   \big|\tilde G^{\la=0}_{1,\tilde L}(i\omega\to \omega + i0)\big|^2\Big]\,,
\end{split}\end{equation}
where $\rho_\tn{lead}(\omega) = -\tn{Im\,}g_\tn{lead}(i\omega\to\omega+i0)/\pi$, and $f(\omega)=1/[1+\exp(\omega/T)]$ is the Fermi function. At finite $T$, there are additional vertex correction to $G$, which, however, vanish within our truncation scheme.

\begin{figure}[t]
\includegraphics[width=0.95\linewidth,clip]{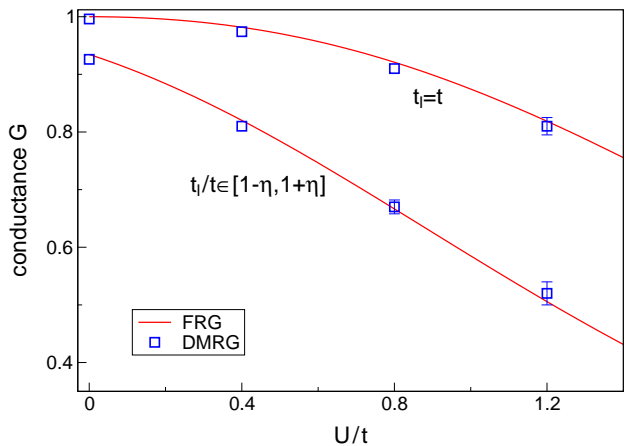}
\caption{(Color online) Zero-temperature conductance of a short quantum wire of $L=8$ sites featuring nearest-neighbor interactions $U$, randomly distributed hoppings $t_l$, and vanishing on-site potentials $V_l=0$. The system is contacted abruptly ($L_c=0$) to tight-binding leads. We compare FRG and DMRG results for the clean case $t_l=t$ as well as for the disorder realization $t_{1\ldots8}/t-1=\{-0.231,-0.153,0.093,-0.253,0.090,0.167,0.047\}$. Error bars for the DMRG data are shown if they are larger than the symbol size (see the main text for details). }
\label{fig:dmrg}
\end{figure}

\subsection{Density Matrix Renormalization Group}

The density matrix renormalization group \cite{white1,dmrgrev} is an algorithm to variationally compute ground states or to simulate the real time evolution \cite{tdmrg1,tdmrg2,tdmrg3,tdmrg4,tdmrg5,tdmrg6} in one-dimensional systems. It can be implemented conveniently using matrix product states.\cite{mps1,mps2,mps3,mps4} Since the DMRG is a fairly standard tool, we only describe briefly of how the conductance is computed;\cite{pcurrent} more details can be found, e.g., in Ref.~\onlinecite{tdmrg4}.

Within the DMRG, we model both the left and right reservoirs in real space as non-interacting tight-binding chains of size $L_\tn{res}=200$. Their hopping amplitudes $t_{l}^\tn{res}$ are chosen constant $t_l^\tn{res}=t^\tn{res}$ close to the contacts with the wire $H_\tn{LL}$ but are decreased exponentially towards the ends in order to reduce finite-size effects. We first apply a bias voltage $\pm V_b/2$ to the left and right chain and determine the ground state of the whole system (limiting ourselves to zero temperature for the DMRG results). Thereafter, we set $V_b=0$ and calculate the real-time evolution of the charge current. We extract the steady-state value for 12 values of $V_b<t^\tn{res}$ and obtain the conductance from linear fits.

The computational effort of DMRG calculations scales with the third power of the dimension of the matrix product state used to approximate a given 1d state, which in turn scales exponentially with the amount of encoded entanglement. Generally speaking, the longer the size of the interacting wire, the more entanglement builds up in the steady state. This limits the DMRG calculations to small values of $\tilde L$. The error is determined by the finite size of the leads, uncertainties in the extracted steady-state current due to oscillations, and finite-entanglement errors (truncation of the matrix product state). We run the DMRG calculation for various parameters (system size, discard weight) in order to roughly estimate the error. For our purposes, it is sufficient that the DMRG data is accurate to a few percent, and we can refrain from a precise analysis of, e.g., finite-time oscillations as discussed in Ref.~\onlinecite{tdmrg4}. Error bars are shown in case that they are larger than the symbol size; we also emphasize that the FRG calculation is by construction exact at $U=0$, implying that this point serves as a benchmark for the DMRG result (see Fig.~\ref{fig:dmrg}).

\begin{figure}[t]
\includegraphics[width=0.95\linewidth,clip]{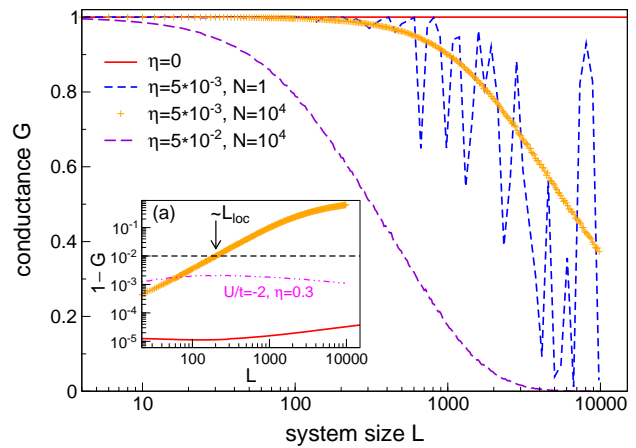}
\caption{(Color online) FRG data for the zero-temperature conductance through large wires of up to $L=10^4$ sites with $U/t=1$, uniform hoppings $t_l=t$, and random on-site disorder potentials of strength $\eta$. The latter are averaged over $N$ different configurations. The system is adiabatically connected to Fermi liquid leads; we take the wide-band limit and choose the hybridization as $\Gamma=t$. Note that only the data for $\eta=0$ as well as for $\eta=0.005, N=10^4$ is shown in the inset. The additional curve for attractive $U$ will be discussed in Sec.~\ref{sec:outlook}. }
\label{fig:g}
\end{figure}

\section{Results}
\label{sec:results}

We first compare our approximate FRG data with the DMRG reference for small systems at zero temperature. Disorder is modelled via random hopping amplitudes in the wire, $t_l/t\in[1-\eta,1+\eta]$, which renders the DMRG calculations simpler since one can trivially stay at half filling.\cite{tdmrg4} We set the on-site potentials $V_l$ to zero, consider non-adiabatic contacts ($L_c=0$), and choose equal hopping strengths $t=\tau=t^\tn{res}$ in the tight-binding leads and in the wire. Results are shown in Fig.~\ref{fig:dmrg} both for the clean system, where the conductance deviates from the unitary values merely due to the abruptness of the contacts, as well as for one disorder realization with $\eta=0.3$. Even though the FRG approximation is a priori justified only to leading order, it agrees well with the DMRG data up to large interactions $U/t\sim1$. We again point out that the FRG calculation is exact at $U=0$ so that this point in turn serves as a non-trivial benchmark for the DMRG result.

We can now use the FRG to study systems as large as $O(10^5)$ sites. In the remainder of the paper, we employ uniform hoppings $t_l=t$ and introduce disorder via random on-site potentials drawn from a uniform distribution
\begin{equation}
V_l/t\in[-\eta,\eta]\,. 
\end{equation}
Furthermore, we choose structureless wide-band limit leads with a hybridization strength of $\Gamma=t$.

Fig.~\ref{fig:g} shows FRG results for the length-dependence of the conductance. For clean systems, $G(L)$ is independent of $L$ and of unitary value if the contacts to the baths are perfectly adiabatic.\cite{localLL1,localLL2,LLcontacts} For our purposes, it is sufficient to choose $L_c=22$, $s=2$ in Eq.~(\ref{eq:usmooth}); this yields $1-G(L=10000)\approx 3\times10^{-8}$ at $U/t=0.2$ and $1-G(L=10000)\approx 3\times10^{-5}$ at $U/t=1$ [see Fig.~\ref{fig:g}(a)]. In presence of a finite $\eta>0$, $G(L)$ is a non-monotonous curve for any given choice of the potentials $V_l$, reflective of randomly distributed transport resonances within the wire. After numerically averaging over $N\sim10^{4}$ different disorder realizations, we obtain a smooth $G(L)$ which decays monotonously for repulsive interactions (for attractive interactions see Sec.~\ref{sec:outlook}). We will now analyze this quantitatively. We will discard all data for which the deviations from the unitary conductance are not at least one order of magnitude larger than the above-mentioned deviations attributed to imperfect contacts.

\begin{figure}[t]
\includegraphics[width=0.95\linewidth,clip]{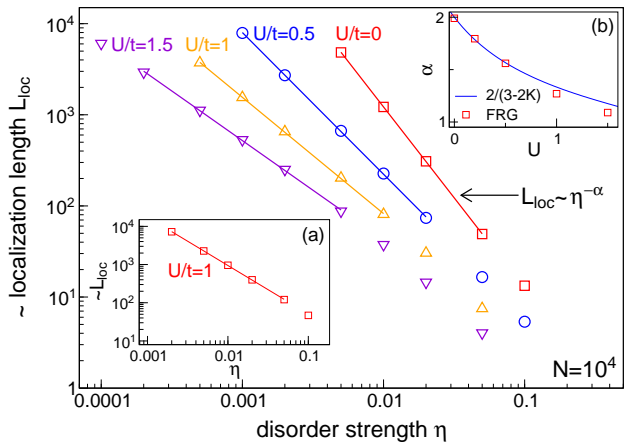}
\caption{(Color online) Localization length $L_\tn{loc}$ as a function of the disorder strength $\eta$ for various interactions $U$. We defined $L_\tn{loc}$ as the length scale where the averaged conductance shown in Fig.~\ref{fig:g} is suppressed to $G(L=L_\tn{loc})=0.99$ [main panel], or $G(L=L_\tn{loc})=0.75$ [inset (a)]. The solid lines show power-law fits to $L_\tn{loc}\sim\eta^{-\alpha}$ for $L_\tn{loc}\gtrsim50$. Inset (b): Exponent $\alpha$ in comparison with the prediction $\alpha=2/(3-2K)$. }
\label{fig:lloc}
\end{figure}

\subsection{Localization length}
\label{sec:lloc}

We define the localization length $L_\tn{loc}$ as the scale on which the zero-temperature conductance starts to deviate from the unitary value [see Fig.~(\ref{fig:g}(a)]. Results are shown in Fig.~\ref{fig:lloc}, where we have used the precise definition $G(L_\tn{loc})=0.99$ in the main panel and $G(L_\tn{loc})=0.75$ in inset (a). The localization length decreases monotonously with the strength $\eta$ of the disorder. For $\eta\sim0.1$, $L_\tn{loc}$ becomes of the order of a few lattice sites. In the limit $\eta\to0$, we observe a power law, $L_\tn{loc} \sim \eta^{-\alpha}$, which is consistent with previous works. \cite{gs1,gs2,eckern} For spinless fermions, an RG analysis within the Tomonaga-Luttinger model \cite{giamarchi} predicts an exponent $\alpha=2/(3-2K)$, where $K$ is the Luttinger liquid parameter, which in our case is known analytically from Bethe ansatz: $K=\pi/\{2\arccos[-U/(2t)]\}$.\cite{haldanebethe} Despite the fact that our FRG data is strictly correct only to leading order in the interaction, it reproduces this result even for large $U/t=1.5$ [see Fig.~\ref{fig:lloc}(b)].

\begin{figure}[t]
\includegraphics[width=0.95\linewidth,clip]{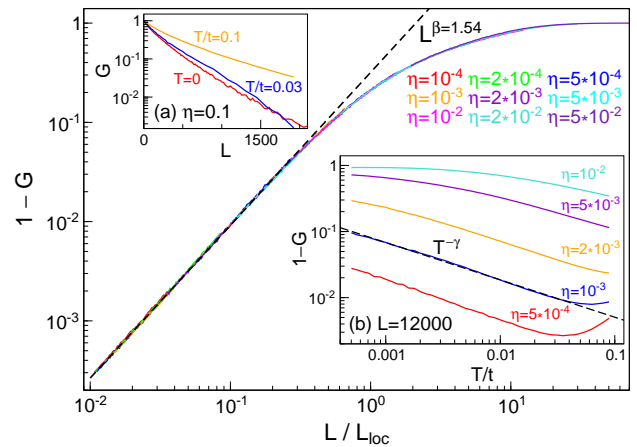}
\caption{(Color online) FRG results for the length- and temperature-dependence of the conductance. \textit{Main panel:} $G(L)$ at $T=0$, $U/t=1$, $200<L<10000$, and various $\eta$. The data was averaged over $N=10^4$ samples. For $L\ll L_\tn{loc}$, we observe power-law behavior $1-G\sim L^{\beta}$ with an exponent $\beta\approx3-2K$ (dashed line). \textit{Inset (a):} $G(L)$ on a log-linear scale for fixed $\eta=0.1$ and different temperatures. The data at $T=0$ and $T>0$ was averaged over $N=10^4$ and $N=10^3$ samples, respectively. \textit{Inset (b):} $G(T)$ for fixed $L=12000$, various $\eta$, and averaged over $N=10^2$ samples. At intermediate $T$, we again observe a power law $1-G\sim T^{-\gamma}$, where $\gamma\approx\beta-1$.}
\label{fig:scaling}
\end{figure}

\subsection{Length and temperature dependence}
\label{sec:scaling}

We now analyze the functional form of the disorder-averaged conductance in more detail. We start at zero temperature. If $L$ is of the order of a few lattice sites, $G(L)$ decays in non-universal way. For $L\gg 1$, however, the data for different values of $\eta$ can be collapsed onto a single curve if $L$ is rescaled w.r.t.~the localization length -- the conductance has a universal form $G(L/L_\tn{loc})$. This is illustrated in Fig.~\ref{fig:scaling}, where we have used $L_\tn{loc}\sim\eta^{-\alpha}$.

For lengths scales $1\ll L\ll L_\tn{loc}$, $G$ decays with a power law, $1-G(L)\sim L^{\beta}$. An analytical guess for the exponent $\beta$ can be obtained from the low-energy analysis of a \textit{spinfull}, homogeneous Luttinger liquid, \cite{apel2} suggesting $\beta=3-2K$ in our case. We find $\beta\approx1.28$ at $U/t=0.5$ and $\beta\approx1.54$ at $U/t=1$ (Fig.~\ref{fig:scaling}), which agrees decently with $3-2K\approx1.277,1.5$. On larger length scales $L\gg L_\tn{loc}$, the conductance $G(L)$ shows an exponential decay, which is still observable at small, finite temperatures [see Fig.~\ref{fig:scaling}(a)]. This is consistent with the system being localized for repulsive interactions at small $\eta$ (see also the discussion in Sec.~\ref{sec:outlook}).\cite{apel1,apel2,apel3,gs1,gs2,eckern}

Finally, we study the behavior of $G$ if the temperature is increased for a fixed value of the system size. Results are shown in Fig.~\ref{fig:scaling}(b). While at small $T$, the conductance is exponentially suppressed, we observe a power-law increase  $1-G(T)\sim T^{-\gamma}$ at intermediate temperatures (larger than the `localization temperature' but smaller than the bandwidth). The exponent is in good agreement with the analytic prediction\cite{giamarchi} $\gamma=2-2K$ obtained for a spinless LL (we find $\gamma\approx0.29$, $2-2K\approx0.28$ at $U/t=0.5$ and $\gamma\approx0.56$, $2-2K=0.5$ at $U/t=1$).

\begin{figure}[t]
\includegraphics[width=0.95\linewidth,clip]{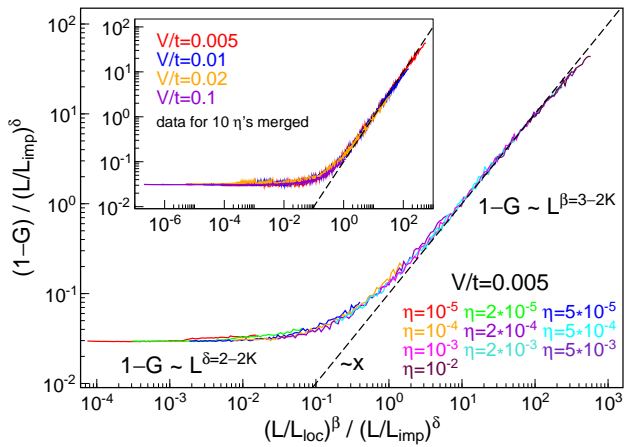}
\caption{(Color online) Scaling of the zero-temperature conductance in presence of on-site disorder $\eta$ as well as a single impurity $V$ at $U/t=0.5$ calculated via the FRG. One observes a crossover between two different power laws; the data for all $V$ and $\eta$ can be collapsed on a single curve. The raw data was obtained for lengths scales $200<L<50000$ and averaged over $N=10^3$ samples. }
\label{fig:imp}
\end{figure}

\subsection{Disorder and isolated impurities}
\label{sec:imp}

One of the hallmarks of Luttinger liquid physics is the influence of isolated impurities.\cite{lutherpeschel,mattis,apel2,gs2,kf} An arbitrarily small barrier
\begin{equation}
H_\tn{imp} = V n_{\tilde L/2}
\end{equation}
effectively cuts a LL in half at low energies, and the conductance vanishes with a power law $G(E\to0)\sim E^{2/K-2}$, where $E$ is an energy scale such as temperature or inverse length. At larger $E$ and small $V$, $G$ approaches the value $G_0$ of the impurity-free Luttinger liquid via $G_0-G(E)\sim E^{2-2K}$. The crossover between those two limits of `weak' and `strong' impurities follows a universal scaling form $G(E/E_\tn{imp})$, where $E_\tn{imp}(V)$ is a $V$-dependent, non-universal scale. This whole picture was first established for homogeneous systems using the local Sine-Gordon model, e.g.~by means of perturbative RG or the Bethe ansatz.\cite{kf,scaling1,scaling1,scaling3} Subsequently, the FRG was used to extensively study impurities in microscopic lattice models,\cite{LLvolker} illustrating that the conductance in presence of leads features the same power laws and single-parameter scaling if and only if the contacts to the leads are adiabatic.\cite{LLvolker2,LLvolker2a,LLvolker3}

If the system features both weak impurities $V$ as well as weak disorder $\eta$, one expects a crossover between the respective power laws $1-G(L)\sim L^{\delta=2-2K}$ and $1-G(L)\sim L^{\beta=3-2K}$. However, it is a priori unclear whether or not the data for different $L$, $V$, and $\eta$ can be collapsed on a universal curve. In order to test this, we make a single-parameter scaling ansatz
\begin{equation}
1 - G(L) = \left(\frac{L}{L_\tn{imp}}\right)^\delta f \left[ \frac{\left(L/L_\tn{loc}\right)^\beta}{\left(L/L_\tn{imp}\right)^\delta}\right]\,,
\end{equation}
where $f$ should satisfy $f(x\to0)\sim1$ and $f(x\to\infty)\sim x$. $L_\tn{loc}$ is the localization length, which for the parameters considered is simply taken as $L_\tn{loc}(\eta)\sim\eta^{-\alpha}$, and we determine $L_\tn{imp}(V)$ as the length scale needed to collapse the data at $\eta=0$ but various $V$ onto a single curve. Results are shown in Fig.~\ref{fig:imp}, indicating that the conductance indeed has a scaling form.

\section{Outlook}
\label{sec:outlook}

We have studied the combined effect of disorder and correlations in a system of 1d lattice fermions using a leading-order functional renormalization group scheme. We computed the conductance in presence of adiabatic coupling to leads on all length and temperature scales. At low energies and for weak disorder, one observes several Luttinger liquid power laws whose exponents are in good agreement with predictions obtained via field theory. The interplay of isolated impurities and disorder is governed by a universal single parameter scaling form.

In this paper we focused on the limit of weak disorder and repulsive interactions where the system is believed to be localized. Prior works suggest that a crossover into a metallic phase occurs at $K=3/2$, which corresponds to attractive $U/t=-1$ in our units. As illustrated by Fig.~\ref{fig:g}(a), the conductance no longer decays to zero at $U/t=-2$ even for large values of $\eta$, and one might be tempted to conclude that our FRG scheme captures the metal-to-insulator transition. Moreover, the persistence of exponentially-decaying $G(L)$ at finite temperature for repulsive interactions and intermediate disorder strengths is indicative of the existence of a many-body localized phase. However, caution is in order: For the weak-disorder limit addressed in this paper, previous works suggest that the interaction gives rise to power laws whose exponents are of first order, and employing a leading-order FRG scheme is thus reasonable. Many-body localization, however, is a strong-correlation phenomenon about which it is not known whether or not it can be captured by leading-order perturbative RG. Hence, it is imperative to implement a true second-order FRG approximation as a frame of reference (which, e.g., accounts for the energy-dependence of the two-particle vertex). This is straightforward for one-dimensional, inhomogeneous systems (see, e.g., Ref.~\onlinecite{frgjan}) and subject of ongoing work. It is certainly not garantueed that a second-order scheme would succeed in describing MBL quantitatively, but one would expect that the MBL phase can be detected and that some of its features can be analyzed qualitatively from the second-order flow towards strong coupling.

\emph{Acknowledgments} --- We thank Volker Meden for his fruitful comments. We acknowledge support by the Nanostructured Thermoelectrics program of LBNL (CK \& JEM), by the AFOSR MURI (JEM), and by the Simons Foundation (JEM).



\begin{thebibliography}{*}
\bibitem{commentflow}
In the flow equation (\ref{eq:flow}) for $U_l^\la$, we also pragmmatically neglect terms proportional to $(G_{l,l+3}^\la)^2$, which are much smaller than the terms $(G_{l,l}^\la)^2$ and $(G_{l,l+1}^\la)^2$.
\end{thebibliography}
\end{document}